\documentclass[twocolumn,superscriptaddress,aps,prb]{revtex4-1}
\usepackage{bm}
\usepackage{amssymb,amsmath}
\usepackage{comment}    
\usepackage{times}
\usepackage{graphicx}
\usepackage{array}
\usepackage{lineno}

\usepackage{float}
\graphicspath{{./graph/}}

\begin{document}

\setpagewiselinenumbers

\title{A Study of Geometry in Anisotropic Quantum Hall States by Principal Component Analysis}
\author{Na Jiang}
\affiliation{Zhejiang Institute of Modern Physics, Zhejiang University, Hangzhou 310027, P. R. China}

\author{Siyao Ke}
\affiliation{Zhejiang Institute of Modern Physics, Zhejiang University, Hangzhou 310027, P. R. China}

\author{Hengxi Ji}
\affiliation{Zhejiang Institute of Modern Physics, Zhejiang University, Hangzhou 310027, P. R. China}

\author{Hao Wang}
\affiliation{Shenzhen Institute for Quantum Science and Engineering and Department of Physics,
Southern University of Science and Technology, Shenzhen 518055, P. R. China}

\author{Zi-Xiang Hu}
\affiliation{Department of Physics and Center for Quantum materials and device, Chongqing University,Chongqing 401331, P. R. China}

\author{Xin Wan}
\affiliation{Zhejiang Institute of Modern Physics, Zhejiang University, Hangzhou 310027, P. R. China}
\affiliation{CAS Center for Excellence in Topological Quantum Computation, University of Chinese Academy of Sciences, Beijing 100190, P. R. China}

\begin{abstract}
In the presence of mass anisotropy, anisotropic interaction, or in-plane magnetic field, quantum Hall droplets can exhibit shape deformation and internal geometrical degree of freedom.
We characterize the geometry of quantum Hall states by principal component analysis, which is a statistical technique that emphasizes variation in a dataset.
We first test the method in an integer quantum Hall droplet with dipole-dipole interaction in disk geometry.
In the subsequent application to fractional quantum Hall systems with anisotropic Coulomb interaction in torus geometry, 
we demonstrate that the principal component analysis can quantify the metric degree of freedom and 
predict the collapse of a $\nu = 1/3$ state.
We also calculate the metric response to interaction anisotropy at filling fractions $\nu = 1/5$ and $2/5$ and 
show that the response is roughly the same within the same Jain sequence, 
but can differ at large anisotropy for different sequences.
\end{abstract}
\date{\today}
\maketitle

\section{Introduction}
\label{sec:introduction}

In recent years quantum Hall states with broken rotational symmetry have been explored experimentally in systems with anisotropic band mass or in the presence of in-plane magnetic field.~\cite{xia11,kamburov13, Lilly99, Jo17}
A key theoretical question is how we can describe these states beyond model wave functions, which only characterize the topological property of the states and which have no apparent variational parameter.
Haldane~\cite{haldane11} pointed out that the geometrical properties of the wave functions have long been overlooked.
The Laughlin wave function, for example, is not necessarily associated with rotational symmetry in Laughlin's original proposal;~\cite{laughlin83} in fact, it represents a family of wave functions, each with its distinct geometrical parameter. 
The wave functions can be constructed explicitly, e.g, via a unimodular transformation, which encodes a global metric tensor to accommodates the intrinsic geometry of the wave functions.~\cite{qiu12} 
These variational wave functions have been demonstrated numerically to characterize the quantum Hall states with mass or interaction anisotropy, as well as in the presence of an in-plane magnetic field.~\cite{qiu12,yang12,wang12,Apalkov14,Papic13,bohu17,YangCPB,HuPRB2018}

An alternative avenue is to explore the wave function responses of the geometrical disturbance, which can be quantified by an anisotropic mass or interaction metric. 
Consider an integer quantum Hall droplet with dipole-dipole interaction.~\cite{qiu11}
The anisotropic interaction effect can be represented by a single mode that distorts the edge of the droplet; the mode has an edge momentum $\Delta M = 2$ for dipolar interaction.~\cite{qiu13}
Similar to the Gutzwiller wave function for the electron correlation in an on-site Hubbard model,~\cite{fuldebook} one can introduce a Jastrow factor to account for the geometrical responses. 
The resulting wave function is consistent with the unimodular construction up to a center-of-mass mode due to the boundary confinement.~\cite{qiu13} 
More exotic excitations, such as emergent gravitons,~\cite{yang12b,yang16, golkar16, yang19} occur in the fractional case, in which the shape of the exchange-correlation hole, in addition to the overall droplet shape, responds to the metric change. 
The emergent FQH graviton can be excited by a geometric quench,~\cite{yang19, liu18,lapa19} which demonstrates its nontrivial dynamics in time. 

On the other hand, large anisotropy in mass or interaction can suppress the fractional quantum Hall state.~\cite{wang12,HuPRB2018} 
Take the Laughlin state at $\nu = 1/3$ as an example, which is a condensate of composite fermions formed by attaching two vortices to each electron. 
Heuristically, we can think of the state as a collection of triple zeros for each electron at the location of other electrons, which is encoded in the Laughlin wave function. 
In the presence of anisotropy, the unimodular transformation split the zeros along the easy axis. 
Even though we can continue to deform the Laughlin state geometrically, the topological order is expected to be broken 
once the split of the triple zeros is comparable to the average distance among electrons. 
The rough estimate of the critical anisotropy that destroys the Laughlin state has been demonstrated numerically 
to be valid. A liquid crystal like phase emerges for larger anisotropy, which is characterized by incommensurate peaks in the projected static structure factor.~\cite{wang12}

Motivated by the rapid development in machine learning,~\cite{goodfellow16} we revisit the anisotropic quantum Hall systems with different interactions and in different geometries. 
Quantum Hall systems with anisotropic interaction are strongly correlated systems, whose difficulties, in particular beyond the description of model wave functions, lie in the high dimension of the Hilbert space. 
We approach with dimension reduction in mind, introducing an unsupervised learning, which does not rely on the existence or knowledge of model wave functions, to explore geometrical information from datasets of many-body wave functions obtained by exact diagonalization. 
Our study finds that the principal component analysis (PCA) method, which has been applied to many physics problems,~\cite{wang16,Car17,Troyer17,Gao17,wang17,wetzel17,hu17,costa17,wang18} serves this purpose well. 
The leading principal component describes the topology of a family of wave functions, while the subleading components describe the geometry of the states. 
In particular, the second principal component allows us to extract geometrical excitations, to quantify the intrinsic metric of wave functions, and to locate the collapse of topological order. 
We find that the PCA study reveals that the geometrical response to anisotropic interaction is roughly unchanged in the same Jain series, but differs in different series. 

The rest of the paper is organized as follows. 
We introduce the models for anisotropic quantum Hall systems and the PCA method in Sec.~\ref{sec:method}. 
The PCA method is tested on integer quantum Hall states on disk geometry with dipole-dipole interaction in Sec.~\ref{sec:IQHE}.
In Sec.~\ref{sec:FQHE} we apply the method to fractional quantum Hall states on torus geometry with anisotropic Coulomb interaction and compare the geometrical responses for filling factor $\nu = 1/3$, $2/5$, and $1/5$.
We summarize our results in Sec.~\ref{sec:summary} and discuss the connection to related work and the potential generalizations. 

\section{Models and Method}
\label{sec:method}

\subsection{Anisotropic Interaction}
\label{models}

In this study we explore models with two types of interaction: dipole-dipole interaction\cite{qiu13} and coulomb interaction with an in-plane dielectric tensor.\cite{wang12} The first interaction is studied in the context of the integer quantum Hall states while the second of the fractional quantum Hall states.

\subsubsection{Dipole-dipole interaction}
\label{model:dipole}
In the Bose-Einstein condensation of $^{52}$Cr atom~\cite{Griesmaier05} or the degenerate quantum gas of $^{40}$K$^{87}$Rb~\cite{Myatt97}, the interaction can be described as dipole-dipole interaction with the $s$-wave
scattering vanishing for spin polarized fermions. Consider the spin-polarized fermionic dipoles in a symmetric potential 
\begin{equation}
U(r) = \frac{1}{2} m (\omega ^2x^2+\omega ^2y^2+\omega _{z}^{2}z^2)
\end{equation} 
with radial trap frequency $\omega$, 
particle mass $m$, and axial trap frequency $\omega_{z}$. 
The system is rotating rapidly around $z$ axis 
with an angular frequency 􏰭$\Omega < \omega$.
In the fast-rotation limit, the system can be regarded as quasi-2D.~\cite{Cooper08, Fetter09, Baranov08, qiu11} 
The motion in the $z$ direction is frozen in its ground state, 
so we need to integrate out the corresponding degree of freedom. 
The wave function of the two-body relative coordinate in this direction is 
\begin{equation}
\phi(z) = \exp(-z^2/4 q^2)/(2 \pi q^2)^{1/4},
\end{equation} 
where $q$ measures the thickness in $z$ direction in units of $l = \sqrt{\hbar/(2 m \omega_z)}$. 
The effective 2D interaction, in the $x$-$y$ plane, has the form
\begin{equation}
  V_{2D}(\vec{\rho}, \theta) = \int dz V_{dd}(\vec{r}, \theta)|\phi(z)|^2
\end{equation}
in units of $d^2/(4\pi \epsilon_0 l^3)$, where
$d$ is the dipole moment and $\epsilon_0$ is the vacuum permittivity.
Here, the polarized interaction is 
\begin{equation}
V_{dd}(\vec{r}, \theta) = \frac{r^2-3(z \cos\theta +x \sin\theta)^2}{r^5},
\end{equation} 
where $\theta$ is the angle between the dipole moment and the $z$ axis. 
At $\theta = 0$, all dipole moment are oriented in $z$ direction, and thus the system has rotational symmetry. 
This symmetry is broken while a nonzero component of dipole moment exists in the $x-y$ plane at $\theta \neq 0$.
 The geometric effect of the anisotropic quantum Hall state and its phase transition can be studied by varying the parameter $\theta$.~\cite{HuPRB2018} For moderate $\theta$, the system remains in the quantum Hall phase without rotational symmetry. A phase transition would be expected for larger $\theta$.   
 
In our PCA study below, we consider the anisotropic IQH regime; in this case, the many-body wave functions for different anisotropy have already be characterized in Ref.~[\onlinecite{qiu13}], which can be directly compared to the PCA results. 

\subsubsection{Anisotropic Coulomb interaction}
\label{model:coulomb}

One can introduce anisotropy into FQH systems with Coulomb interaction through anisotropic dielectric tensor or anisotropic band mass, both of which can be represented by a set of generalized pseudopotentials.~\cite{bohu17} In this study we explore the geometrical effect in several FQHE states with anisotropy Coulomb interaction with the form
\begin{equation}
  V(\rho, A_{c}) = \frac{e^2}{4\pi \epsilon_{0} \sqrt{g^{ab}\rho_{a} \rho_{b}}}
\end{equation}
where $g^{ab}$ is diagonal:
\begin{equation}
  g^{ab} = \begin{pmatrix}
  1/A_c & 0 \\
  0     & A_c \\
\end{pmatrix}
= \begin{pmatrix}
  e^{-A} & 0 \\
  0     & e^{A} \\
\end{pmatrix}
\end{equation}
with $A = \ln A_{c}$. For strong enough magnetic field, the Hamiltonian can be projected into the LLL. 
The form of the projected Hamiltonian on the torus with Landau gauge is~\cite{rezayi00}
\begin{equation}
  H=\frac{1}{N_{\phi}}\sum_{q}V(\vec{q})e^{-q^2/2}\sum_{i<j}e^{i\vec{q} \cdot (\vec{R_{i}}-\vec{R_{j}})}
\end{equation}
where $N_{\phi}$ is the total quantum flux through the rectangular unit cell and $\vec{R_{i}}$ is the guiding center coordinate of the $i$th electron. 
The Fourier transform of the anisotropic Coulomb interaction is 
\begin{equation}
V(\vec{q}) = 1/\sqrt{q_x^2/A_c + A_c q_y^2}.
\end{equation}
The momentum components $q_{x}$ and $q_{y}$ are integral multiples of $2\pi / L_{x}$ and $2\pi / L_{y}$, respectively.
For moderate anisotropy, the system remains in the FQH phase, while larger $A_{c}$ can drive the system into a liquid-crystal-like phase.~\cite{wang12} We study the geometrical effect due to the interaction anisotropy by the PCA for various filling fractions, including 1/3 and 2/5 in the first Jain sequence 
and $1/5$ in the second Jain sequence.~\cite{Jainbook}

\subsection{Principal Component Analysis}
\label{model:pca}

Quantum Hall systems, as well as other many-body systems, have a huge Hilbert space. 
In most case, however, we are only interested in the ground state and a few low-lying excited states. 
This means that dimension reduction can play an important role in understanding many-particle physics. 
In this respect, modern machine learning methods play 
a similar role in extracting limited features from a large dataset of complex systems. 
These methods explore the fact that even though we have a huge amount of data 
with considerably many features, the majority of these features that can be used 
to describe the situation are correlated with each other, 
leading to much smaller dimensions of interest. 
Methods of dimension reduction are, therefore, crucial in better understanding 
the complex systems. 

One of the widely used dimension reduction techniques is the PCA.~\cite{pearson01}
PCA reduces the dimension of samples by linearly projecting them onto 
a new feature space of fewer dimensions. 
These new features are called the principal components, which
are the main directions along which samples distribute. 
By using principal components to describe the samples, 
one can find out their characteristics efficiently.

In this study, we consider a family of normalized real wave functions with a parameter 
that describes the guiding center geometry of the states. 
Suppose the wave functions are represented by 
\begin{equation}
\left \vert \psi^{(\alpha)} \right \rangle = \sum_{i=1}^{D} c^{(\alpha)}_i \left \vert i \right \rangle,
\end{equation}
where $\left \vert i \right \rangle$ represents the many-particle basis with dimension $D$, 
while $\alpha$ labels the set of $M$ ground state wave functions 
with different anisotropy parameter. 
The PCA searches for a projection matrix $P$ with dimension $d \times D$ 
that best reproduces the wave functions in reduced dimensions; in other words,
\begin{equation}
\sum_{\alpha = 1}^M \left \Vert P^T P \left \vert \psi^{(\alpha)} \right \rangle 
- \left \vert \psi^{(\alpha)} \right \rangle \right \Vert^2
\end{equation}
is minimized. 
Technically, we set up a data matrix of the following form
\begin{equation}
  X =
  \begin{pmatrix}
    c^{(1)}_1 & c^{(1)}_2 & \cdots & c^{(1)}_D \\
    c^{(2)}_1 & c^{(2)}_2 & \cdots & c^{(2)}_D \\
    \vdots & \vdots & \ddots & \vdots \\
    c^{(M)}_1 & c^{(M)}_2 & \cdots & c^{(M)}_D \\
    \end{pmatrix}.
\end{equation}
The projection is carried out toward the subspace spanned by the 
eigenvectors of the covariance matrix $X^T X$ with the largest 
$d$ eigenvalues. 
In practice, the PCA can find these eigenvectors by the 
singular value decomposition $X=U\Sigma V^T$ of the data matrix $X$,
where $U$ and $V$ are orthogonal matrices and $\Sigma$ a diagonal matrix. 
The covariance matrix of $X$ is thus
\begin{equation}
\label{eq:pca1}
X^TX = V \Sigma^2 V^T.
\end{equation}
$V^T$ transform the matrix of wave functions $X$ to 
\begin{equation}
Y = XV = 
  \begin{pmatrix}
    y^{(1)}_1 & y^{(1)}_2 & \cdots & y^{(1)}_D \\
    y^{(2)}_1 & y^{(2)}_2 & \cdots & y^{(2)}_D \\
    \vdots & \vdots & \ddots & \vdots \\
    y^{(M)}_1 & y^{(M)}_2 & \cdots & y^{(M)}_D \\
    \end{pmatrix}.
\end{equation}
The projection is performed in the sense that we are only interested 
in the first $d$ diagonal elements of $\Sigma$ 
and the first $d$ columns of $y$. 
The resulting matrix element $y^{(\alpha)}_i$ is, therefore, the projected 
amplitude of the $\alpha$th wave function along the $i$th principal axis. 
We note that the covariance matrix of $Y$ is diagonal
\begin{equation}
\label{eq:pca2}
  Y^TY = \Sigma^2 \equiv M {\rm diag}(\lambda_1, \lambda_2, \cdots, \lambda_D),
\end{equation}
where $\lambda_i$s are known as explained variance ratios and satisfy
$\sum_i \lambda_i = 1$. 
The principal components with large explained variance ratios
spanned the subspace that is an approximate representation of 
the original set of wave functions. 
Our goal is to use the resulting projected amplitudes in this subspace 
to quantify the guiding center geometry of  the wave functions. 

\section{IQH States on Disk Geometry}
\label{sec:IQHE}

In this section, we apply PCA to study the wave function deformation of the
IQH state in disk geometry with dipole-dipole interaction.
The goal here is to demonstrate the feasibility and the simplicity of PCA
in understanding the geometrical information of a family of wave functions of
the same topological character.
As discussed earlier, PCA emphasizes variation and brings out dominating
features in a dataset.
The analysis in this example thus decipher topology from geometry.

For concreteness, we consider the microscopic system with dipole-dipole
interaction as discussed in Sec.~\ref{model:dipole}.
In a strong harmonic trap with $\alpha = 1.0$, the dipolar fermions are
confined at its maximum density except at the perimeter of the
droplet.~\cite{qiu11}
For continuously varying polar angle of the dipoles, we obtain a family of IQH
wave functions
\begin{equation}
\label{eq:SMA approx}
\Psi_{\gamma} = e^{ -\gamma \sum_{i<j}(z_i-z_j)^2}
\left [ \prod_{i < j} (z_i - z_j) \right ]
e^{-\sum_i |z_i|^2/4},
\end{equation}
where $\gamma$ is a variational parameter that describes the geometric shape or
the deformation of the IQH droplet.~\cite{qiu13}
Due to the nontrivial interaction, these states are not simply product states.
Their wave functions can be described by the product of the isotropic IQH state
and a Jastrow factor that arises in the single-mode approximation of a model
quadrupolar interparticle interaction
$V(i,j) \propto \Re (z_i - z_j)^2$.~\cite{qiu13}

The variational wave function suggests that the anisotropic IQH ground state
can be written as the superposition of the isotropic IQH state and
its edge states with angular momentum increment of integral multiples of 2.
These edge states are of the form
\begin{equation}
\label{eq:SMA components}
\Phi_2^p = N_p \left [\sum_{i<j}(z_i-z_j)^2
  \right ]^p \prod_{i<j}(z_i-z_j) e^{-\sum_i|z_i|^2/4},
\end{equation}
where $p$ is a positive integer and $N_p$ the normalization factor.~\cite{qiu13}
The edge states are orthogonal to each other in the Hilbert space and are,
thus, expected to be the principal components, up to a unitary transformation.

\begin{figure}
  \centering
  \includegraphics[width=\linewidth]{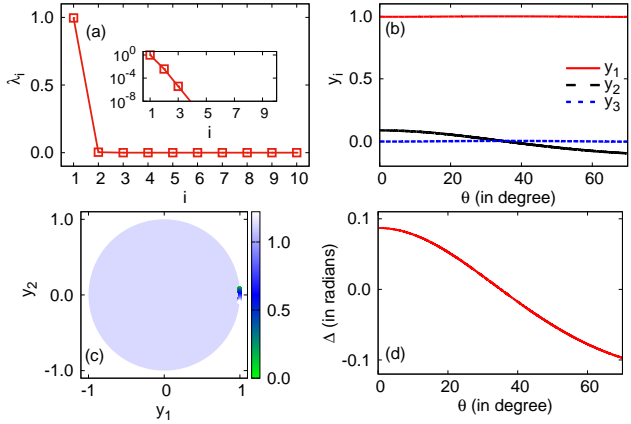}
  \caption{(Color online) PCA results and analysis for the IQH state
    with dipole-dipole interaction.
    (a) The first few explained variance ratios
    obtained from raw ground state wave functions. The inset displays with log
    scale. (b) Projections of the various anisotropy
strength $\theta$ to the three leading principal components. (c) Projection of
the samples onto the plane of the two leading principal components. The
colorbar indicates the anisotropy strength. (d) The angle $\Delta$ between the
first projection and the second one as a function of anisotropy strength.}
  \label{fig:iqh results}
\end{figure}

We feed PCA with ground state wave functions for M = 701 polar angles $\theta$ 
distributed uniformly between zero and $70^{\circ}$.
Fig.~\ref{fig:iqh results}(a) shows the largest 10 explained variance ratios, 
among which the first three are sufficiently dominant. 
This means that even though the interacting IQH states live in a high-dimensional space, 
their evolution can be well approximated in a rather low-dimensional space.  
The projected amplitudes $y_1$, $y_2$, and $y_3$ along the three principal axes 
are shown in Fig.~\ref{fig:iqh results}(b).
The deviation of $y_3$ from zero is already difficult to see by naked eyes,
so the evolution is a two-dimensional rotation in the lowest, but an excellent, approximation.  
We confirm the two-dimensional evolution by plotting $y_1$ versus $y_2$ 
in Fig.~\ref{fig:iqh results}(c), in which the data falls on the perimeter of the unit circle. 
The data clusters near $x$-axis, suggesting that the rotation is limited. 
By plotting the polar angle $\Delta$  (in radians) of the right side of the data in Fig.~\ref{fig:iqh results}(c) 
against the polar angle $\theta$ (in degrees) of the dipoles in Fig.~\ref{fig:iqh results}(d),
we obtain a geometrical characterization of the ground state wave functions 
for various $\theta$. 

\begin{figure}
  \centering
  \includegraphics[width=\linewidth]{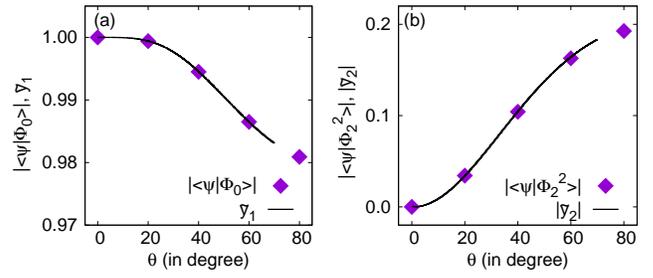}
  \caption{(Color online) Comparison between rotated projected amplitudes and overlaps of wave functions with $\Phi_0$ and $\Phi_2^2$ in Ref.[9]. (a) The first rotated projected amplitudes $\tilde{y}_1$(black solid) is consistent with the overlaps(square) of wave functions with $\Phi_0$. (b) The absolute value of second rotated projected amplitudes $\tilde{y}_2$(black solid) consist with the overlaps(square) of wave functions with $\Phi_2^2$.
  }
  \label{fig:compare with Qiu}
\end{figure}

The PCA results of the ground state evolution are expected to be consistent with 
the wave function decomposition into $\Phi_2^p$ [Eq.~(\ref{eq:SMA components})] 
based on the physical ground. 
The quantitative agreement needs an extra rotation because $\Delta_0 \equiv \Delta (\theta = 0) \neq 0$.
By minimizing fluctuations, PCA selects the ground state with $\theta = 35^{\circ}$ 
to be the first principal component, as evident in Fig.~\ref{fig:iqh results}(c). 
After a two-dimensional rotation,
\begin{equation}
\label{eq:rotation}
\left ( \begin{array}{c} 
\tilde{y}_1 \\
\tilde{y}_2
\end{array} \right ) = 
\left [ \begin{array}{cc} 
\cos(\Delta_0) & \sin(\Delta_0) \\
-\sin(\Delta_0) & \cos(\Delta_0)
\end{array} \right ]
\left (
\begin{array}{c} 
y_1 \\
y_2
\end{array} \right ),
\end{equation}
we expect the first component becomes the isotropic IQH state $\Phi_0$ 
and the second $\Phi_2^2$. 
Fig.~\ref{fig:compare with Qiu} compares $\tilde{y}_1$ and $\tilde{y}_2$ 
with the overlaps of wave functions with $\Phi_0$  and $\Phi_2^2$, respectively, 
which have been calculated in Ref.~[\onlinecite{qiu13}]. 
The excellent agreement confirms that PCA, as a well-established tool for dimensional reduction,  
is effective in separating the geometrical evolution (in subleading principal components) 
from topology (in the leading component). 

\begin{figure}
  \centering
  \includegraphics[width=\linewidth]{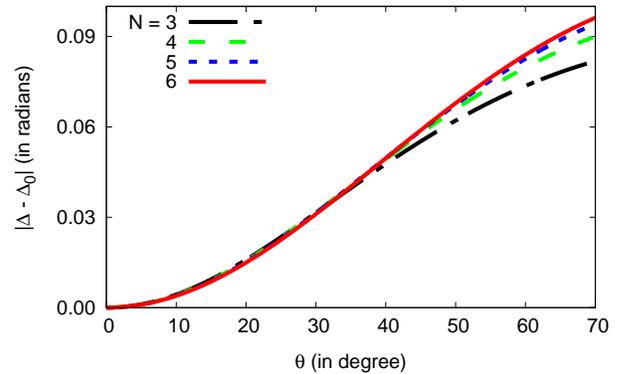}
  \caption{(Color online)
The absolute value of $\vert \Delta - \Delta_0 \vert$ (in radians) as a function of $\theta$ (in degree) for systems with $N = 3-6$ particles. For larger systems, the angle curves collapse onto one.
  }
  \label{fig:various Nel}
  \end{figure}

For small geometrical distortion, $\tilde{y}_2$ can be obtained by $\Delta - \Delta_0$ 
without carrying out the explicit rotation in the reduced space. 
Therefore, the quantitative comparison suggests that we can identify $y_2$, 
the projection of the ground state on the subleading principal component, 
with $\gamma$, the metric parameter in the variational wave function [Eq.~(\ref{eq:SMA approx})], 
up to a rotation. 
Fig.~\ref{fig:various Nel} plots $\vert \Delta - \Delta_0 \vert$ as a function of $\theta$
for  systems with $N = 3$-6 particles. 
The system size dependence is found to be negligible for $\theta < 40^{\circ}$, 
where the geometrical distortion is sufficiently small so higher-order contributions 
(beyond two leading principal components) can be omitted. 

\section{FQH States on Torus Geometry}
\label{sec:FQHE}

The application of PCA to the anisotropic IQH system is a vivid
demonstration of the statistical learning method.
The necessity of the method in disk geometry is debatable, as there exist
versatile approaches, such as the Jack polynomial diagonalization~\cite{lee14}
and the Monte Carlo algorithm,~\cite{zhang14} to relate the wave
functions in the first and the second quantization forms.
In torus geometry, however, we have less tools.
It is, therefore, an interesting problem to explore the applicability of the
PCA in closed, translationally invariant systems,
especially for the cases that the ground states are not in the form of
model wave functions that are exact solutions of corresponding model
Hamiltonians.

For this purpose, we turn to the model with anisotropic Coulomb interaction
as introduced in Sec.~\ref{model:coulomb}.
In the torus geometry, the ground state wave functions, hence the data matrices, are complex.
The transpose in Eqs.~(\ref{eq:pca1}) and (\ref{eq:pca2}), therefore, 
needs to be replaced by conjugate transpose. 
The projected amplitude $y_i = y_i^r + i y_i^i$ is now complex.  

Earlier study for $\nu = 1/3$ filling~\cite{wang12} has showed that the
Laughlin state remains to be stable but anisotropic for weak interaction anisotropy.
For sufficiently strong anisotropy, the system undergoes a transition from the
FQH liquid to a liquid-crystal-like state.~\cite{wang12}
Therefore, the motivation of using the PCA here are two-fold.
First, for a large range of anisotropy, can the PCA identify the phase
transition between the competing ground states?
Second, in the Laughlin phase, can the PCA quantify the guiding center metric
of the FQH wave functions?

We focus on three families with $\nu = 1/3$, $2/5$, and $1/5$ in the following.
The first two families belong to the same Jain sequence, in which two flux
quanta are attached to each electron in the composite fermion construction.
The third, however, combines four flux quanta to each electron in the flux
attachment, hence can have more complex geometrical responses.

\subsection{$\nu = 1/3$}
\label{state:laughlin3}

\begin{figure}
  \centering
  \includegraphics[width=\linewidth]{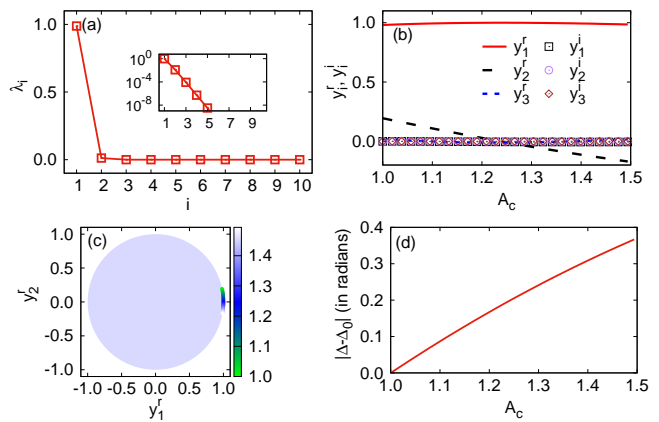}
  \caption{(Color online)
PCA results for anisotropic Coulomb interaction at $\nu = 1/3$ for $N=10$ electrons with anisotropy strength from 1.0 to 1.5. (a) The largest 10 explained variance ratios on linear scale and insert exponential scale. (b) The real and imaginary part of the projected amplitudes $y_1^{r(i)}, y_2^{r(i)}, y_3^{r(i)}$ for the first three components. (c) The projections of the samples onto the plane of $y_1^r$ and $y_2^r$. The colorbar indicates the anisotropy strength $A_c$. (d) The absolute value of the shifted polar angle $\vert \Delta - \Delta_0 \vert$ (in radians) as a function of anisotropy strength $A_c$.}
  \label{fig:1/3 results}
  \end{figure}

We consider electrons with anisotropic Coulomb interaction at $\nu = 1/3$ and 
feed PCA with ground state wave functions for $N=10$ electrons with different interaction anisotropy 
from $A_c = 1.0$ to 1.5.
Fig.~\ref{fig:1/3 results}(a) shows the largest 10 explained variance ratios
on both linear and exponential scales. 
Compared with the integer case, the second ratio becomes visibly nonzero on the linear scale, 
suggesting that the effect of anisotropy is stronger in this range of $A_c$ in the fractional case. 
Fig.~\ref{fig:1/3 results}(b) shows the real and imaginary part of the projected 
amplitudes for the first three components. 
Even though the wave functions are complex, 
to a good approximation the imaginary part of the projected 
amplitudes can be neglected, which means that the geometrical effect of 
the Laughlin state can be roughly described by a real representation. 
Compared to the two leading components, $y_3 = y_3^r + i y_3^i$ and other higher-order terms 
can still be neglected.  
As shown in Fig.~\ref{fig:1/3 results}(c), the pair $(y_1^r, y_2^r)$ of the data falls on the perimeter 
of the unit circle.
This, again, allows us to calibrate the geometry of the wave function by 
the shifted polar angle $\vert \Delta - \Delta_0 \vert$ in Fig.~\ref{fig:1/3 results}(d).

\begin{figure}
  \centering
  \includegraphics[width=\linewidth]{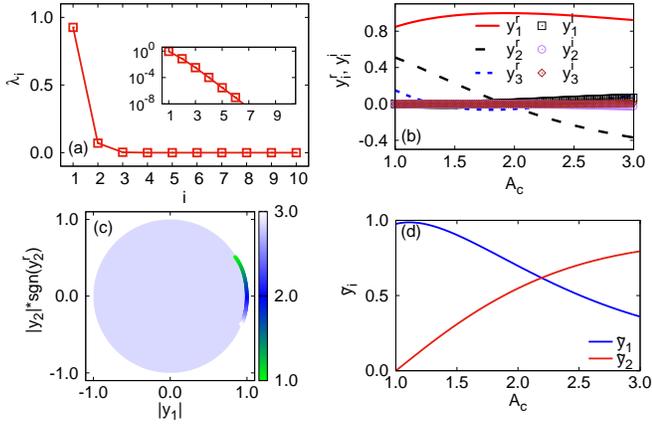}
  \caption{(Color online) PCA results for anisotropic Coulomb interaction at $\nu=1/3$ for $N=10$ electrons with anisotropy strength from 1.0 to 3.0. (a) The largest 10 explained variance ratios on linear scale and insert exponential scale. (b) The real and imaginary part of the projected amplitudes $y_1^{r(i)}, y_2^{r(i)}, y_3^{r(i)}$ for the first three components. (c) The projections of the samples onto the plane of $\vert y_1 \vert$ and ${\rm sgn}(y_2^r) \vert y_2 \vert$. The colorbar indicates the anisotropy strength $A_c$. (d) The rotated projected amplitudes $\tilde{y}_1$ and $\tilde{y}_2$ as functions of anisotropy strength $A_c$.}
  \label{fig:1/3 collapse}
 \end{figure}

Next, we analyze 106 different ground state wave functions for $A_c = 1.0-3.0$. 
The second largest explained variance ratio is now visibly nonzero, 
as shown in Fig.~\ref{fig:1/3 collapse}(a). 
Fig.~\ref{fig:1/3 collapse}(b) shows that the imaginary part of the projected 
amplitudes is still negligible, compared with the corresponding real part. 
The isotropic ground state has projected components 0.85, 0.51, and 0.15 along the 
three leading principal directions. 
We can neglect $y_3$ again, which contributes no more than 2.3\% to the ground states, 
and plot $\vert y_1 \vert$ versus ${\rm sgn}(y_2^r) \vert y_2 \vert$ in Fig.~\ref{fig:1/3 collapse}(c).  
To a good approximation, the data falls on the perimeter of the unit circle, 
indicating that the evolution can be described by the relative weight change 
of two wave functions. 
After a rotation in the $\vert y_1 \vert$-${\rm sgn}(y_2^r) \vert y_2 \vert$ plane, 
as defined in Eq.~(\ref{eq:rotation}), 
we plot $\tilde{y}_1$ and $\tilde{y}_2$ as functions of $A_c$ in Fig.~\ref{fig:1/3 collapse}(d).  
As discussed in the IQH case, $\tilde{y}_1$ is the projection of the ground state 
wave functions on the isotropic one at $A_c = 1$. 
The projection $\tilde{y}_2$ on the other axis exceeds $\tilde{y}_1$ at $A_c = 2.2$, 
indicating a phase transition from the Laughlin phase to a different phase 
induced by strong anisotropic interaction. 
This value is in good agreement with $A_c \approx 2.0$ identified by the sudden collapse of 
the excitation energy gap.  
In practice, the rotation is not necessary, because the transition point can be 
determined by $\vert \Delta_c - \Delta_0 \vert = \pi / 4$.

\subsection{$\nu = 2/5$}
\label{state:jain}

\begin{figure}
  \centering
  \includegraphics[width=\linewidth]{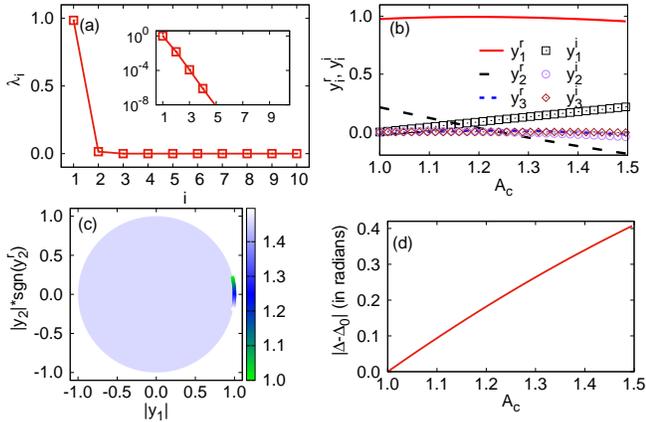}
  \caption{(Color online) PCA results for anisotropic Coulomb interaction at $\nu = 2/5$ for $N=10$ electrons with anisotropy strength from 1.0 to 1.5. (a) The largest 10 explained variance ratios on linear scale and insert exponential scale. (b) The real and imaginary part of the projected amplitudes $y_1^{r(i)}, y_2^{r(i)}, y_3^{r(i)}$ for the first three components. (c) The projections of the samples onto the plane of $\vert y_1 \vert$ and ${\rm sgn}(y_2^r) \vert y_2 \vert$. The colorbar indicates the anisotropy strength $A_c$. (d) The absolute value of the shifted polar angle $\vert \Delta - \Delta_0 \vert$ (in radians) as a function of anisotropy strength $A_c$.}
  \label{fig:2/5 results}
  \end{figure}

The FQH effect at $\nu = 1/3$ can be regarded as the $\nu = 1$ IQH effect of 
composite fermions, in which two magnetic flux quanta are attached to each electron.  
We now turn to an $N = 10$ electron system at $\nu = 2/5$ in the same Jain sequence 
with two composite fermion LLs filled. 
We apply PCA to 32 ground state wave functions from anisotropic Coulomb interaction
with $A_c = 1.0-1.5$. 
Fig.~\ref{fig:2/5 results} shows the 10 leading explained variance ratios, 
the amplitudes projected to the first three axes, 
the evolution of $(\vert y_1 \vert, {\rm sgn}(y_2^r)\vert y_2 \vert)$,
and the variation of $|\Delta - \Delta_0|$ as a function of $A_c$. 
The results are very similar to those in Fig.~\ref{fig:1/3 results}, 
implying that the additional composite fermion LL does not affect the metric of the 
wave functions. 
We note that the imaginary part of $y_i$ can, again, be neglected. 

\begin{figure}
  \centering
  \includegraphics[width=\linewidth]{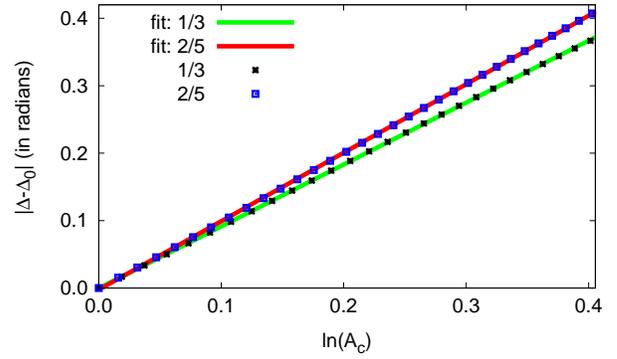}
  \caption{(Color online)
     Comparison of $\vert \Delta - \Delta_0 \vert$ between $\nu = 1/3$ and $2/5$ states. The data can be fitted by a straight line $0.92\ln A_c$ for $1/3$ case and $1.02\ln A_c$ for $2/5$ state.}
  \label{fig:fit 1/3 2/5}
  \end{figure}

To quantitatively compare the geometrical effect for $\nu = 1/3$ and 2/5,
we plot $\vert \Delta - \Delta_0 \vert$ as a function of $A = \ln A_c$ for the two cases in 
Fig.~\ref{fig:fit 1/3 2/5}. 
Data in each case can be fitted by a straight line
\begin{equation}
\label{eq:linear fit}
\vert \Delta - \Delta_0 \vert = c \ln A_c 
\end{equation}
where the slope $c = 0.92$ for $\nu = 1/3$ and $c = 1.02$ for $\nu = 2/5$.  
The linear dependence in Eq.~(\ref{eq:linear fit}) can be understood as the linear response 
of the wave function metric to the interaction metric, as there is only one parameter 
in the wave functions, which also characterizes the split of the two flux quanta 
from each electron in the composite fermion picture. 
The 10\% difference in the prefactor of the linear term is likely due to the fact 
that each composite fermions LL at $\nu = 2/5$ has too few particles, 
because noticeable deviations also show in the IQH case for $N = 3$ and 4 electrons 
in Fig.~\ref{fig:various Nel}. 

\subsection{$\nu = 1/5$}
\label{state:laughlin5}

\begin{figure}
  \centering
  \includegraphics[width=\linewidth]{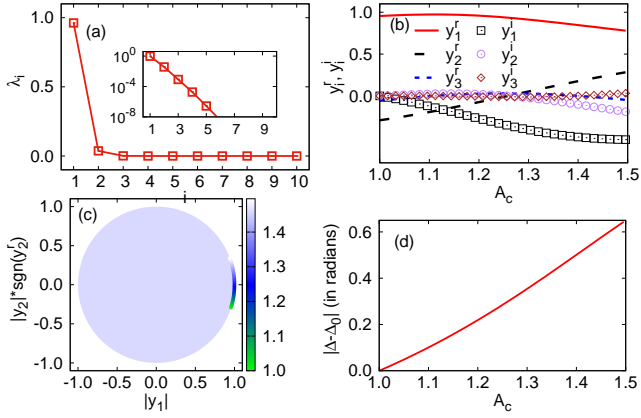}
  \caption{(Color online) PCA results for anisotropic Coulomb interaction at $\nu = 1/5$ for $N=8$ electrons with anisotropy strength from 1.0 to 1.5. (a) The largest 10 explained variance ratios on linear scale and insert exponential scale. (b) The real and imaginary part of the projected amplitudes $y_1^{r(i)}$, $y_2^{r(i)}$, $y_3^{r(i)}$ for the first three components. (c) The projections of the samples onto the plane of $\vert y_1 \vert$ and ${\rm sgn}(y_2^r) \vert y_2 \vert$. The colorbar indicates the anisotropy strength $A_c$. (d) The absolute value of the shifted polar angle $\vert \Delta - \Delta_0 \vert$ (in radians) as a function of anisotropy strength $A_c$.}
  \label{fig:1/5 results}
  \end{figure}

The similarities between $\nu = 1/3$ and $2/5$ motivate us to explore the comparison 
between $\nu = 1/3$ and $1/5$. 
The latter two correspond to the filling of $\nu = 1$ composite fermion LL. 
However, for $\nu = 1/5$, there are four flux quanta attached to each electron. 
In the presence of geometrical distortion, it is not obvious why the four flux quanta 
should split in proportion, hence nonlinear dependence in $\ln A_c$ can go 
beyond Eq.~(\ref{eq:linear fit}). 
We consider 8 electrons at $1/5$ filling with anisotropic Coulomb interaction 
for $A_c = 1.0-1.5$ in torus geometry. 
The PCA results for 32 ground state wave functions are summarized 
in Fig.~\ref{fig:1/5 results}. 
Unlike in Fig.~\ref{fig:1/3 results}(b), we find that the projected amplitudes now have significant 
imaginary parts. 
However, the magnitudes of the first two components still dominate and, again, 
fall roughly on the perimeter of the unit circle. 
Nevertheless, the resulting variation of $|\Delta - \Delta_0|$ bends up as 
a function of $A_c$ in Fig.~\ref{fig:1/5 results}(d), as oppose to the bending down 
in Fig.~\ref{fig:1/3 results}(d).
This indicates that the geometrical effect is stronger in the $1/5$ case. 

\begin{figure}
  \centering
  \includegraphics[width=\linewidth]{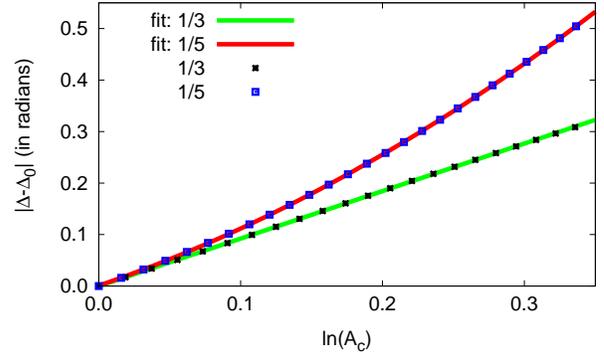}
  \caption{(Color online)
    Comparison of $\vert \Delta - \Delta_0 \vert$ between $\nu = 1/3$ and $1/5$ states. The data can be fitted by $0.92\ln A_c$ for $1/3$ case and $0.95\ln A_c+1.64(\ln A_c)^2$ for $1/5$ state.}
      \label{fig:fit 1/3 1/5}
  \end{figure}

For further comparison, we plot $|\Delta - \Delta_0|$ 
as a function of $A = \ln A_c$ in Fig.~\ref{fig:fit 1/3 1/5} 
for both $\nu = 1/3$ and 1/5. 
We choose the same range of $A_c$ from 1.0 to 1.4 to compare and 
note that the dependence on the range is negligible, 
as long as we do not approach the critical $A_c$ for the collapse of 
the FQH states.
In particular, strong anisotropy also destroy the 1/5 state, 
but the critical $A_c$ is estimated by PCA to be 1.62. 
For $\nu = 1/3$, we find 
\begin{equation}
\vert \Delta - \Delta_0 \vert_{\nu = 1/3} = 0.92 \ln A_c,
\end{equation}
as discussed above. 
For $\nu = 1/5$, on the other hand, the curve can be fitted by 
\begin{equation}
\vert \Delta - \Delta_0 \vert_{\nu = 1/5} = 0.95 \ln A_c + 1.64 (\ln A_c)^2.
\end{equation}
Interestingly, the linear responses are roughly equal in the two cases, 
which is also not far from 1.02 for $\nu = 2/5$, 
while the $1/5$ case has an additional quadratic contribution 
that cannot be neglected. 

\section{Summary and Discussion}
\label{sec:summary}

In this paper, we propose to use PCA, a popular statistical learning method, 
to study the geometrical responses of the quantum Hall wave functions to 
anisotropic interaction. We demonstrated that for moderate anisotropy, 
the emphasis on variation by PCA allows a natural separation of 
topology and geometry. 
The analysis quantifies the geometrical effect in the projection to the axis 
corresponding to second largest explained variance ratio, 
while the leading one encodes the topological wave function, 
up to a trivial rotation. 

The method can also quantitatively identify the transition from 
the topological phase to the CDW phase induced by anisotropy. 
Interestingly, PCA reveals that the wave function evolution 
with increasing anisotropic Coulomb interaction can be 
approximated by a linear interpolation of two wave functions, 
one representing the topological phase and the other the CDW phase.  
This approximation is satisfactory all the way across the anisotropy 
induced quantum phase transition. 

For FQH states in the primary Jain sequence, in which 
each composite fermion contains two magnetic flux quanta, 
PCA finds that the geometrical responses are linear in 
$(\ln A_c)$, which is the logarithm of the diagonal element 
in the anisotropic interaction metric. The amplitude of the 
responses are comparable for $\nu = 1/3$ and $2/5$, 
which have one and two filled composite fermion LLs, 
respectively. 

On the other hand, for a different Jain sequence in which 
each composite fermion has four magnetic flux quanta, 
PCA reveals a strong nonlinear geometrical responses. 
The geometrical effect for $\nu = 1/5$, quantified by the projection to 
the subleading axis, is dominantly quadratic in $(\ln A_c)$, 
even though it has a linear contribution of a similar amplitude 
as in the $1/3$ case. 
The surprising result suggests that the four flux quanta 
in each composite fermions are not split in simple linear fashion. 
Further wave function analysis of the $\nu = 1/5$ case 
in disk or sphere geometry is needed for a clearer picture. 

In a very recent paper, Ippoliti {\it et al.}~\cite{ippoliti18} studied the geometry of 
flux attachment in anisotropic FQH states 
with anisotropic mass and isotropic Coulomb interaction, 
which is equivalent to isotropic mass with anisotropic interaction 
after introducing anisotropic LL orbitals. 
The authors used an infinite density matrix renormalization group (iDMRG) algorithm 
to study the response of the internal wave function metric to band mass anisotropy, 
which was extracted from guiding center structure factor. 
They found that the geometrical response is approximately the same for states 
in the same Jain sequence, but differs substantially between different sequences. 
For $\nu = 1/3$, we draw a similar conclusion that the geometrical 
response is dominated by a linear term, which corresponds to 
the internal unimodular metric.
However, for $\nu = 1/5$, while the iDMRG study found 
significant difference in the prefactor of the linear response 
from that of $\nu = 1/3$, 
we find a similar linear response but very different quadratic response. 
We note that the iDMRG study also found larger quadratic responses 
in isotropic rescaling, but the results showed strong size dependence. 

The main advantage of the PCA in this study, 
compared to more conventional method,~\cite{qiu12} 
is that one can quantify geometrical degree of freedom without the explicit 
knowledge of model wave functions. We thus expect the approach 
can be easily generalized to more complex filling fractions, where 
explicit wave functions cannot be given analytically, or are difficult 
to represent numerically. 
In addition, systems with disorder can also be treated 
with this technique. 

\section{Acknowledgements}
\label{sec:acknowledgements}
The work at Zhejiang University was supported 
by the National Natural Science Foundation of China through Grant No. 11674282, 
the Strategic Priority Research Program of Chinese Academy of Sciences Grant No. XDB28000000,
and the National Basic Research Program of China through Project No. 2015CB921101.
HW is supported by the National Natural Science Foundation of China Grant No. 11474144.
ZXH is supported by the National Natural Science Foundation of China Grants No. 11674041 and No. 11974064.

\end{document}